\documentclass[conference]{IEEEtran}
\IEEEoverridecommandlockouts
\usepackage{amsmath,amssymb,amsfonts}
\usepackage{algorithm}
\usepackage{algpseudocode}
\usepackage{graphicx}
\usepackage{textcomp}
\usepackage{booktabs}
\usepackage{array}
\usepackage{xcolor}
\usepackage{makecell}
\usepackage{listings}
\usepackage{multirow}
\lstdefinestyle{mylistingstyle}{
  basicstyle=\ttfamily\footnotesize,     
  breaklines=true,
  aboveskip=15pt, 
  breakindent=2em,
  breakatwhitespace=true,
  frame=lines,                    
  rulecolor=\color{black},         
  belowcaptionskip=1\baselineskip,
  showstringspaces=false,         
  basicstyle=\footnotesize\ttfamily,
}
\usepackage{tikz}
\usepackage{microtype}
\usetikzlibrary{arrows, positioning, shapes.geometric}
\usepackage[numbers]{natbib}
\def\BibTeX{{\rm B\kern-.05em{\sc i\kern-.025em b}\kern-.08em
    T\kern-.1667em\lower.7ex\hbox{E}\kern-.125emX}}
\begin{document}

\title{Distributed Temporal Graph Learning with Provenance for APT Detection in Supply Chains}

\author{\IEEEauthorblockN{Zhuoran Tan}
\IEEEauthorblockA{\textit{University of Glasgow, UK} \\
z.tan.1@research.gla.ac.uk}
\and
\IEEEauthorblockN{Christos Anagnostopoulos}
\IEEEauthorblockA{\textit{University of Glasgow, UK} \\
Christos.Anagnostopoulos@glasgow.ac.uk}
\and
\IEEEauthorblockN{Jeremy Singer}
\IEEEauthorblockA{\textit{University of Glasgow, UK} \\
jeremy.singer@glasgow.ac.uk} 
}

\maketitle

\begin{abstract}

Cyber supply chain, encompassing digital asserts, software, hardware, has become an essential component of modern Information and Communications Technology (ICT) provisioning. However, the growing inter-dependencies have introduced numerous attack vectors, making supply chains a prime target for exploitation. In particular, advanced persistent threats (APTs) frequently leverage supply chain vulnerabilities (SCVs) as entry points, benefiting from their inherent stealth. Current defense strategies primarly focus on prevention through blockchain for integrity assurance or detection using plain-text source code analysis in open-source software (OSS). However, these approaches overlook scenarios where source code is unavailable and fail to address detection and defense during runtime. To bridge this gap, we propose a novel approach that integrates multi-source data, constructs a comprehensive dynamic provenance graph, and detects APT behavior in real time using temporal graph learning. Given the lack of tailored datasets in both industry and academia, we also aim to simulate a custom dataset by replaying real-world supply chain exploits with multi-source monitoring.
\end{abstract}

\section{Introduction and Motivation}

Since the Solarwind attack report in 2020~\cite{solarwinds}, which impacted major enterprises and millions of end users, there has been a growing tend of exploiting third-party supply chain software to launch attacks, including ransomware \cite{mandianttrends2024} and APTs \cite{europeanunionagencyforcybersecurity.ENISAThreatLandscape2021}. The exploitation has also extended to the AI supply chain, which encompasses pre-trained models, dataset, and dependent libraries \cite{jiangEmpiricalStudyArtifacts2022}.
To counter these threats, most existing solutions focus on detection at the source code level or adopt a prevention-oriented approach. Common methods include identifying malicious code or functions through program analysis \cite{pingyan_wang__2024}, graph mining \cite{10.1145/3691621.3694950}, and machine learning or deep learning techniques \cite{LU2024112031}. Prevention strategies often leverage blockchain frameworks \cite{ISLAM2022100505} to introduce an additional authentication layer for detecting potential tampering. 
However, these approaches assume that third-party source code is readily available for evaluation and benchmarking. In practice, this is rarely the case due to intellectual property protections, limiting the applicability of existing defense mechanisms.

Current work on APT detection primarily relies on provenance-based frameworks, where system behaviors are represented as dynamic graphs \cite{postuvan_learning-based_2024} composed of entities extracted from source data, such as audit logs. These methods are typically designed for a single data source and evaluated using standardized benchmark datasets \cite{edq8-nk52-21} that reflect general APT activities. However, exploitation through SCVs follows a distinct exploitation chain, which may not be effectively captured by detection methods designed for general APTs.

To address current challenges, we propose the following research questions (RQs):
\paragraph{\textbf{RQ1}} What are the unique characteristics of APTs exploiting SCVs??
\paragraph{\textbf{RQ2}} How can such APTs be accurately detected within large-scale source data?
\paragraph{\textbf{RQ3}} How can real-time and efficient detection be achieved while continuously adapting to emerging threats?

\section{Background and Related Work}

APTs exploiting SCVs leverage the supply chain as an entry point and follow a multi-stage attack strategy, attributes, making them more stealthy and difficult to identify in the early stages, thereby maximizing their impact.
SCVs have been observed targeting widely used open-source software repositories such as PyPi, NPM, and Ruby. Additionally, the increasing reliance on pre-trained models hosted on platforms like Hugging Face has introduced new attack vectors within the AI supply chain. Once compromised, these libraries and models can propagate malicious payloads downstream, potentially affecting millions of end users and devices.

Detection efforts for SCVs primarily focus on identifying vulnerabilities at the source code level. Tran et al. \cite{tran_detectvul_2024} proposed a BERT-based architecture that leverages self-attention mechanism to detect vulnerabilities in Python code. Lu et al. \cite{LU2024112031} integrated graph structural information and in-context learning to enhance large language model (LLM)-based vulnerability detection, in which the framework identifies relevant code examples based on semantic, lexical and syntactics similarities. However, these approaches have a significant limitation: they assume access to source code, which is not always available in real-world scenarios.

\begin{figure*}[htbp]
    \centering
    \includegraphics[scale=0.8]{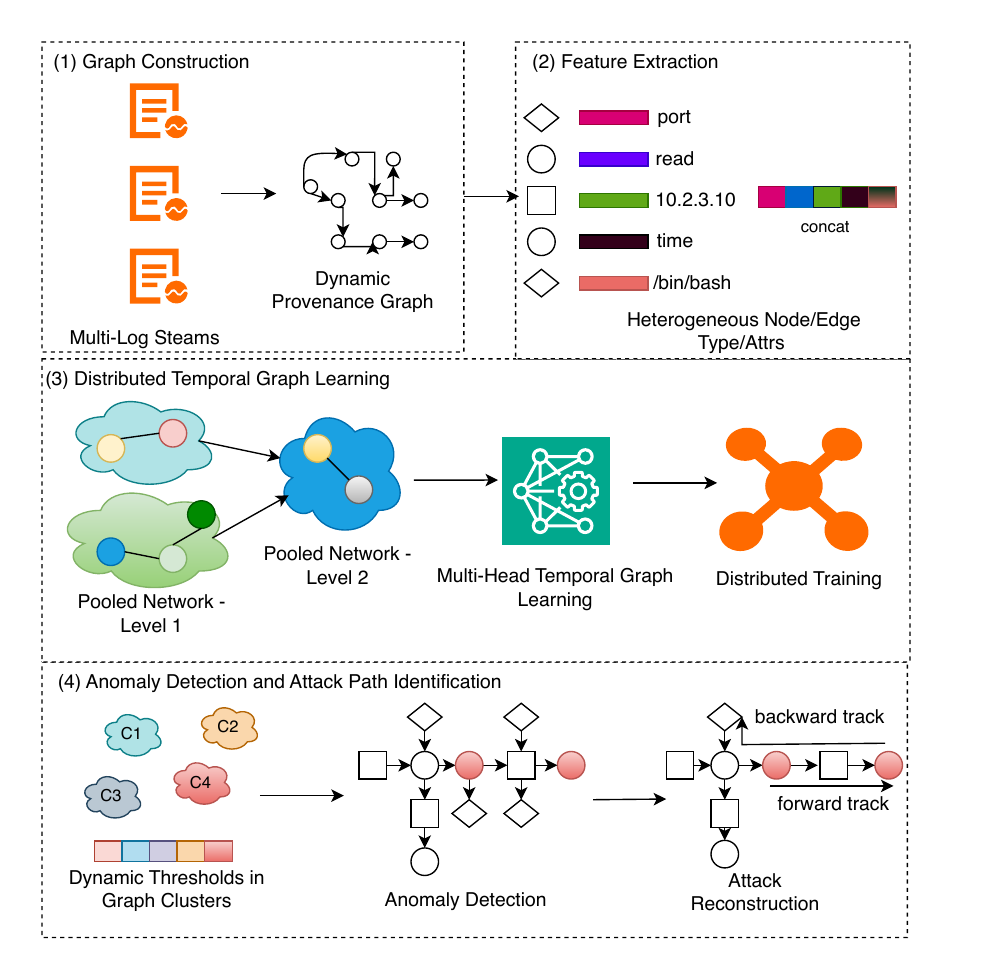}
    \caption{Detection Framework for APTs exploiting Supply Chain Vulnerabilities}
    \label{fig:framework}
\end{figure*}

APT detection focuses on identifying exploitation chains by correlating indicators from multiple data sources, aiming to reconstruct attack paths and trace them back to their root causes \cite{jian2025}. Among existing approaches, data provenance has been widely adopted to correlate timeline-based source data and capture system activity \cite{ xiongPracticalRealTimeAPT2022,Alsaheel2021ATLASAS}, particularly in distributed setting ups \cite{ametepeDataProvenanceCollection2021}. However, the datasets commonly used in these studies do not effectively represent supply chain vulnerability exploitation. Additionally, most detection methods rely on a single type of data source, which is often insufficient for identifying complex and sophisticated APTs.

\section{Methodology and Experiments}

To address \textbf{RQ1}, we conducted a comprehensive survey \cite{10838587} that includes a statistical analysis comparing techniques used in general APT detection and those tailored for APTs exploiting SCVs. The findings reveal the unique exploitation chain of APTs in SCVs.

For \textbf{RQ2}, we first simulated a dataset capturing general OSS exploitation behavior at runtime \cite{DBLP:journals/corr/abs-2411-14829}. This dataset includes 9,461 reports across ecosystems such as npm, PyPI, crates.io, NuGet, and Packagist, providing insights into runtime exploitation by analyzing dynamic behavioral patterns. Additionally, we are simulating a second dataset that models real-world advanced supply chain exploitations on Azure Cloud. This dataset focuses on multi-stage exploitation beyond the initial trigger, addressing the gap in available datasets for studying APTs in SCVs.

To process multi-source data into dynamic, comprehensive provenance graphs, we developed UTLParser \cite{tan2024unifiedsemanticlogparsing}, a scalable tool designed to parse diverse structured data into temporal provenance graphs. These graphs serve as the foundation for subsequent detection methods.

Our current detection approach follows two directions. The first leverages attention-based graph neural networks to extract critical features from the dataset \cite{DBLP:journals/corr/abs-2411-14829}. The second explores the use of temporal graph models for precise attack detection and reconstruction, utilizing the second dataset. To enhance detection accuracy, we implement dynamic threshold-based scoring to identify and reconstruct the most critical attack paths, effectively reducing false positives.

For \textbf{RQ3}, we plan to implement distributed learning techniques to accelerate sampling within the graph dataset and efficiently allocate training tasks across multiple clusters or machines. Additionally, during graph construction, we aim to prune unnecessary structures to optimize message passing and reduce computational overhead.

Regarding continual model updates, we proposed integrating techniques like Elastic Weight Consolidation (EWC) into the continual learning process \cite{Tan2024Ledlog}. This method mitigates catastrophic forgetting during model updates, enabling adaptation to newly identified threats.

As shown in Figure \ref{fig:framework}, our method comprises four key components: graph construction, feature extraction, distributed temporal graph learning, and anomaly detection with attack reconstruction. This framework effectively addresses the identified challenges and overcomes the limitations of existing solutions.
    
\section{Evaluation and Discussion}

To evaluate the performance of our detection framework, we will utilize two simulated datasets focusing on SCVs and APTs based on SCVs. Additionally, we will conduct benchmark testing using a common dataset \cite{edq8-nk52-21}. To further assess practical performance, we plan to implement red teaming with advanced exploitation techniques, evaluating detection efficiency and accuracy in real-world scenarios.

We anticipate challenges related to real-time graph construction and distributed training due to the large volume of data. To mitigate storage and computational overhead, we plan to apply clustering and pruning techniques to filter out redundant information and reduce graph size while preserving essential structures for effective detection.


\end{document}